\documentclass[graybox]{svmult}

\usepackage{type1cm}        
\usepackage{makeidx}         
\usepackage{graphicx}        
\usepackage{multicol}        
\usepackage[bottom]{footmisc}

\usepackage{newtxtext}       %
\usepackage[varvw]{newtxmath}       

\usepackage{hyperref}
\usepackage{subfig}

\newcommand\arcsec{\hbox{$^{\prime\prime}$}}

\newcommand{\simon}[1]{#1}

\usepackage{orcidlink}

\makeindex             

\begin{document}

\title*{Dynamical Modelling of Galactic Kinematics using Neural Networks}
\author{David A. Simon \orcidlink{0000-0001-5742-2982} and\\ Michele Cappellari \orcidlink{0000-0002-1283-8420} and\\
Shude Mao \orcidlink{0000-0001-8317-2788} and\\
Jiani Chu \orcidlink{0009-0008-6513-0427} and\\
Dandan Xu}
\institute{David A. Simon \at Sub-department of Astrophysics, Department of Physics, University of Oxford, Denys Wilkinson Building, Keble Road, Oxford OX1 3RH, \email{david.simon@physics.ox.ac.uk}
\and Michele Cappellari \at Sub-department of Astrophysics, Department of Physics, University of Oxford, Denys Wilkinson Building, Keble Road, Oxford OX1 3RH
\and Shude Mao \at Department of Astronomy, Tsinghua University, Beijing, 100084, People's Republic of China\\
Department of Astronomy, School of Science, Westlake University, Hangzhou, Zhejiang 310030, People's Republic of China
\and Jiani Chu \at Department of Astronomy, Tsinghua University, Beijing, 100084, People's Republic of China
\and Dandan Xu \at Department of Astronomy, Tsinghua University, Beijing, 100084, People's Republic of China}
%
%
\maketitle

\abstract{The advent of integral field data has revolutionised the study of galaxy evolution. A key component of this is dynamical modelling methods which have allowed for crucial insights to be made from kinematic data. Despite this importance, most dynamical models make a number of key assumptions which do not hold for real galaxies. These include assumptions about the geometry (axisymmetry or triaxiality), the shape of the velocity ellipsoid, and the shape of the underlying stellar distribution. At the same time, machine learning methods are becoming increasingly powerful, with many applications appearing in astronomy. \simon{As a first step towards building} new dynamical modelling methods with machine learning, it is important to understand the types of machine learning architectures that are best fit for dynamical modelling. To investigate this, we construct a training set of dynamical models of early-type galaxies using Jeans Anisotropic Modelling (JAM). We then train a neural network on this data using the parameters of JAM and mock photometry as the input. We are able to accurately model JAM galaxies with relatively simple machine learning architectures, leading to a significant speed increase over traditional JAM modelling.}

\section{Introduction}
Dynamical modelling codes make a number of assumptions which are well motivated but don't entirely hold for real galaxies. These include assumptions on the shape of the stellar distribution (often described using a multi-gaussian expansion), assumptions on the shape of the velocity ellipsoid (spherical/cylindrical for Jeans modeling), and assumptions on the underlying geometry of the stellar system (spherical, axisymmetric, or triaxial). \simon{As a first step towards creating} a machine learning dynamical modelling method which breaks these assumptions, we study a machine learning implementation of the code JAM  \cite{cappellari2008jam} to study which neural network architectures are able to describe kinematic data. \simon{JAM models the second moment of the line of sight velocity distribution, a key quantity for studying galaxy dynamics, which is then compared to observations to evaluate a likelihood.}

\section{Machine Learning Model}
\subsection{Training Data}
Following \cite{gomer2023ml} we construct our training data by creating JAM \cite{cappellari2008jam} models of galaxies with sersic photometry. We show our parameters in \autoref{tab:traindat}. In order to train the machine learning architecture more effectively, it is important to only sample parameters corresponding to physical solutions to the Jeans equation. With this in mind, we add an additional constraint to our parameter sampling from \cite{wang2021beta}. There, they show that galaxies with $\beta_z > 0.7 \epsilon_{\rm intr}$ correspond to non-physical solutions of the Jeans equation.
\begin{table}
	\centering
	\caption{Randomly sampled parameters used to generate the training data. Parameters are sampled according to a uniform distribution within the bounds.}
	\begin{tabular}{ |c|c|c| } 
		\hline
		Parameter & Description & Bounds \\ 
		\hline
		$n_{\rm ser}$ & Sersic Index & [2, \ 4] \\ 
		$R_{\rm ser}$ & Sersic Radius & [5\arcsec, \ 20\arcsec] \\ 
		$q_{\rm L}$ & Light Axial Ratio & [0.6, \ 1] \\ 
		$q_{\rm M}$ & Mass Axial Ratio & [0.6, \ 1] \\ 
		$\beta_{\rm z}$ & Anisotropy & [-0.4, \ 0.4] \\ 
		$i$ & Inclination & [$\arccos(0.6)$, \ 90$^\circ$] \\
		\hline
	\end{tabular}
	\label{tab:traindat}
\end{table}

One key difference between this work and \cite{gomer2023ml} is that we do not directly feed all of the parameters in \autoref{tab:traindat} into our machine learning architecture. Rather, we use the parameters $n_{\rm ser}$, $R_{\rm ser}$, and $q_{\rm L}$ to generate mock photometry of the sersic galaxies which is then fed into the machine learning algorithm. The dimensions of the images are $64\times 64$. 

After randomly sampling the parameters above we generate the corresponding JAM models. To do this, we first generate an MGE corresponding to each set of mock-photometry. This is done using the routine \textsc{mge\_fit\_1d} from the \textsc{mgefit} package of \cite{cappellari2002mge} to the sersic profile with axial ratio equal to 1, and then setting the axial ratio for each MGE element equal to $q_{\rm L}$. We then use the routine \textsc{jam\_axi\_proj} from \cite{cappellari2008jam} under the assumption of cylindrical symmetry to generate the velocity maps. One key difference between this work and \cite{gomer2023ml} is that rather than sample the kinematic maps on a regular square grid, we sample the maps on a grid uniformly spaced in $\theta$ and logarithmically spaced in radius. We do this partly to improve the speed of the architecture since it dramatically reduces the number of data points output by the algorithm, but also because JAM internally operates by evaluating the Jeans equation on this $\theta$-log radius grid and then interpolates to the desired points. Thus, sampling on this grid is well motivated from the JAM method, is faster than sampling on a square grid, and can be readily applied to real observations since it is straight forward to interpolate on this grid to whatever data points are required by observations. Due to axisymmetry, we only sample these points on one quarter of the field of view.

We use a single cpu core to generate 10000 samples for the training data. This takes approximately 3 hours. This is significantly faster than the time it took to generate the training data in \cite{gomer2023ml}. 

\subsection{Machine Learning Architecture}
\begin{figure}
	\centering
	\subfloat{\includegraphics[width = \columnwidth]{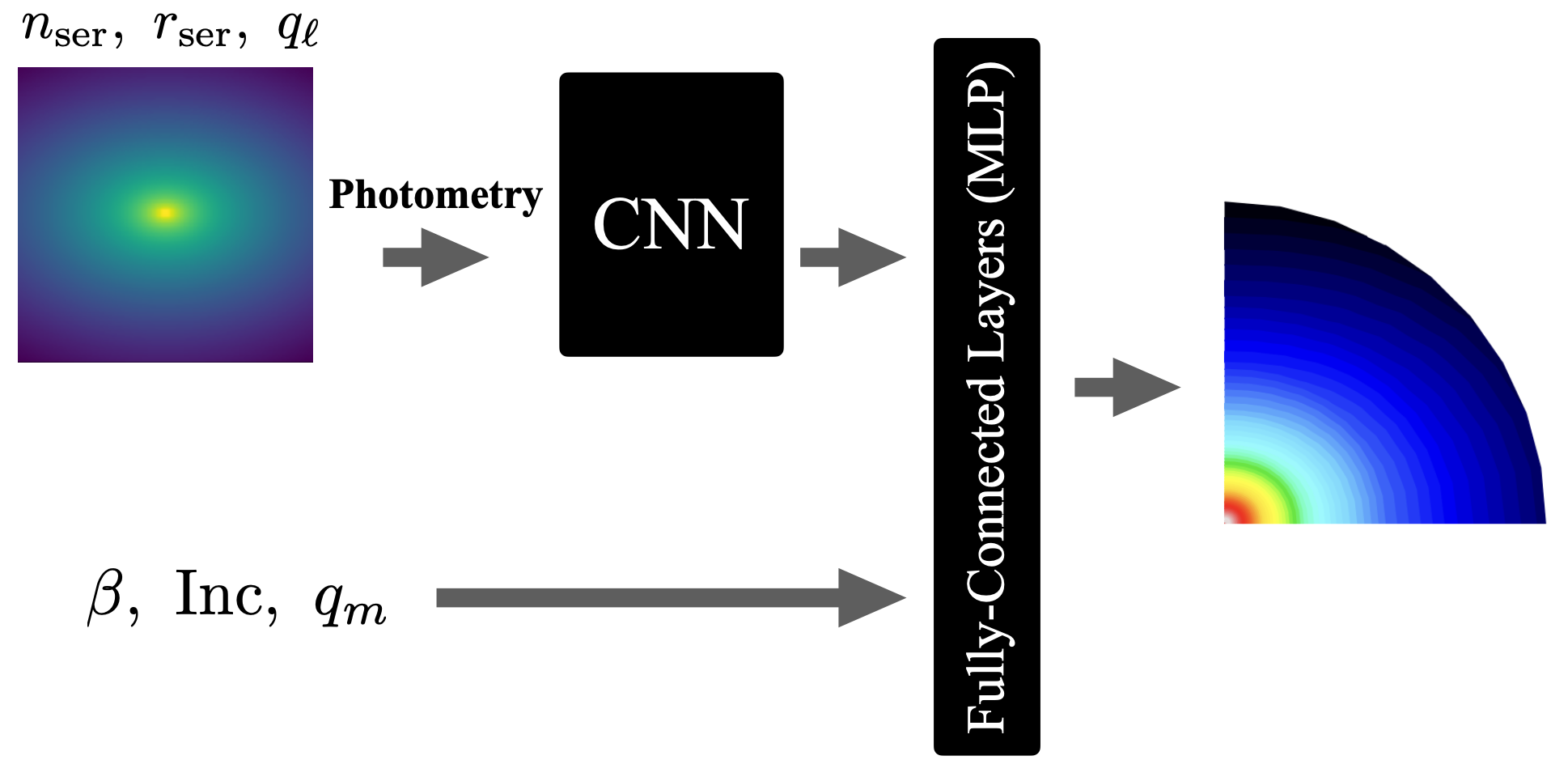}}
	\caption{A schematic of our neural network architecture. Note that due to axisymmetry, the output only features data from one quarter of the total field of view.}
	\label{fig:ml_arch}
\end{figure}

We show a diagram of our neural network architecture in \autoref{fig:ml_arch}. For the first part of the machine learning architecture, we use a convolutional neural network (CNN) to transform the mock photometry into a feature vector of length 64. We experimented with the size of the feature vector and found diminishing returns beyond 64. The network has a total of 6 layers: four of these are convolutional layers with hidden layer size of 128 and 2 of these are 2 by 2 max pool layers. The convolution layers have kernel size 2 and a stride of 2. 

For the second section, we use a 7 layer MLP with input size of 67 (3 for the non-photometry input parameters of \autoref{tab:traindat} and 64 from the output of the CNN), hidden layer size of 4096, and output size of 200 to match the grid of the JAM training data. We chose LeakyReLU with a slope of 0.01 as our activation function. \simon{Note that this architecture exactly copies the work flow of JAM. Another valid choice would have been to directly feed in the second moment maps and photometry and extract the physical parameters. In the standard JAM picture these parameters would be extracted from observations using something like MCMC. Note that our architecture for reconstructing the second moment maps does not use convolutions or anything else which explicitly takes into account information from neighboring grid points. In principle this would increase the accuracy of the model (at the cost of the performance). However, as we will see in \autoref{sec:accuracy}, the current architecture achieves sufficient accuracy.}

\subsection{Training}
 \simon{We train with the mean squared error loss function} using the ADAM optimizer with an initial learning rate of 0.001. To allow for a dynamically updated learning rate we use the PyTorch scheduler ReduceLROnPlateau. This works by decreasing the learning rate by a factor $\lambda$ after the loss has failed to decrease after a certain number of epochs called patience. We set the patience to 20 and the factor to $10^{-1/8}$. We use a batch size of 50 and train for  800 epochs. The training galaxies are reshuffled at the start of each epoch. We use 8000 galaxies for the training set and retain 2000 for the test set. Our training is done using a single NVIDIA V100 GPU. The time it takes to train is approximately 30 minutes.

\section{Performance}
\subsection{Accuracy}\label{sec:accuracy}
\begin{figure}
	\centering
	\subfloat{\includegraphics[width = \columnwidth]{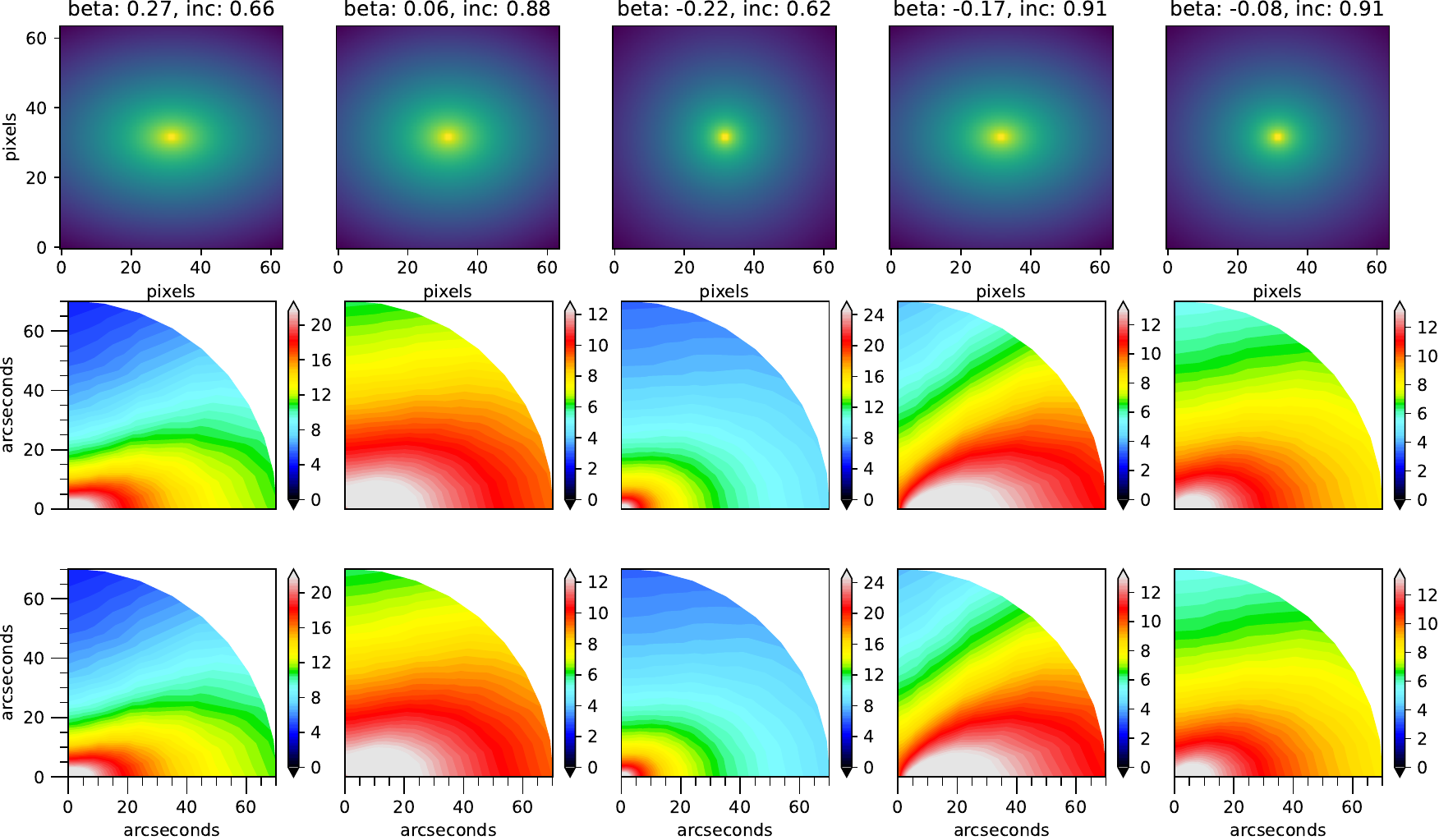}}
	\caption{Random sample of test set galaxies and their performance. Top row: Mock photometry used for a test set data point with kinematic parameters $\beta$ and inclination. Center row: True JAM kinematics. Units are normalized. Bottom row: recovered ML prediction from the photometry and kinematic parameters. The maps are virtually indistinguishable by eye.}
	\label{fig:ml_test_set}
\end{figure}

In principle, a machine learning model can be trained to arbitrary accuracy given sufficient training data and model complexity. Real astrophysical observations have an accuracy of 6-7 km s$^{-1}$ \cite{cappellari2011atlas}, so this is the benchmark accuracy one should aim for. \simon{Any machine learning model which achieves an accuracy significantly lower than this is unnecessarily trading performance for accuracy.} We show some outcomes of our model for the test set in \autoref{fig:ml_test_set}. The top row shows the mock photometry, the central row the true JAM model, and the bottom row the ML prediction. We see that they are identical by eye. 

\begin{figure}
	\centering
	\subfloat{\includegraphics[width = \columnwidth]{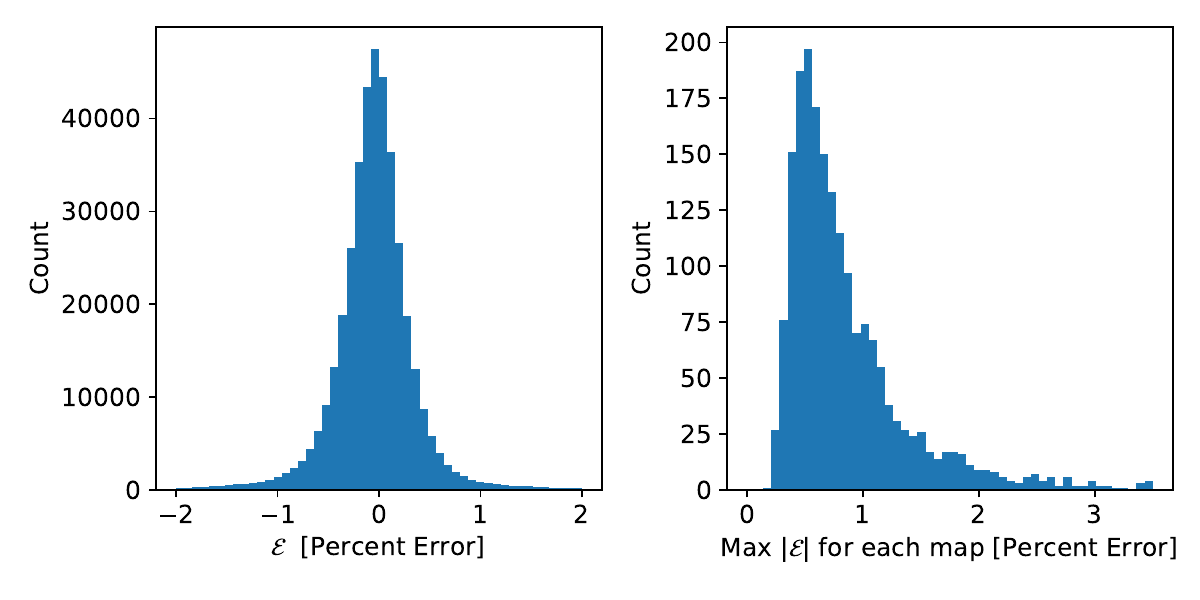}}
	\caption{Left panel shows a histogram of the percent error of every pixel for every galaxy in the test set of galaxy kinematics. Almost all of the errors are less than 1 per-cent. Right panel shows a histogram of the maximum error for each galaxy in the test data. Here we see that the max error for almost every galaxy is less than 3 per-cent.}
	\label{fig:err_hist}
\end{figure}

We quantify the accuracy of our model as $\mathcal{E} = \text{(Model - Truth)}/\text{Truth}$. We show the results for this for our test data in \autoref{fig:err_hist}. In the top panel we show a histogram of the results for every data point in the test data. In the bottom panel we show the results for this where we only consider the maximum of $|\mathcal{E}|$ over each $V_{\rm RMS}$ map in the test data. From this we see that the vast majority of the maps are accurate to within 1 per-cent and 90 per-cent of the maps have a max-error less than 2 per-cent. The maximum $V_{\rm RMS}$ one is likely to encounter in most galaxies is close to 100-300 km s$^{-1}$, so 2 per-cent accuracy corresponds with an error of 2-6 km s$^{-1}$. It is worth noting that this \simon{is not as accurate as} that of \cite{gomer2023ml} which trains using fewer galaxies, but uses a significantly more complicated machine learning architecture and does not use mock photometry. \simon{This hints that there are many potential viable machine learning architectures that can be used for dynamical modelling which can be selected on the basis of whether or not one is trying to optimize for performance, training set size, or accuracy.}

\subsection{Speed}
In addition to accuracy, it is important to carefully benchmark the speed of the model. Because the machine learning architecture we use is a simple convolutional neural network fed into a multi-layer perceptron, we expect significant speed improvements over previous work. To quantify this, we randomly sample 1000 input parameters and record the time it takes to run the model for each set of parameters. We find that it takes on average 1 milli-second to run the model. This, however, over-estimates the total amount of time required to run the model. This is because the CNN part of our machine learning architecture only needs to be evaluated one time for each choice of photometry. Thus, for practical uses, such as running MCMC chains, it is only necessary to run the CNN for each set of photometry one time, and then one can repeatedly vary parameters in the MLP. As the majority of the time spent evaluating the model is spent in the CNN, the total time to run the model is closer to 0.3 milli-seconds. This is faster than the model in \cite{gomer2023ml} by around a factor of 300.

\section{Conclusion}
This work has demonstrated that it is possible to use simple machine learning architectures to accurately perform dynamical modelling. In the future we hope to extend this work to more general classes of training data. This could potentially allow us to break some of the constraints enforced by current dynamical models, allowing for more accurate dynamical modelling of galaxies.

%
%
%

\end{document}